\documentclass[prd,twocolumn]{revtex4}
\usepackage{graphicx}
\usepackage{amssymb}
\usepackage{epstopdf}

\begin{document}

\title{Stable, accelerating universes in modified-gravity theories}
\author{Simon DeDeo}
\affiliation{Kavli Institute for Cosmological Physics, University of Chicago, Chicago, IL 60637}
\author{Dimitrios Psaltis}
\affiliation{Department of Physics, University of Arizona, Tucson, AZ 85721}
\date{\today}

\begin{abstract}
Modifications to gravity that add additional functions of the Ricci
curvature to the Einstein-Hilbert action -- collectively known as
$f(R)$ theories -- have been studied in great detail. When considered
as complete theories of gravity they can generate non-perturbative
deviations from the general relativistic predictions in the solar
system, and the simplest models show instabilites on cosmological
scales. Here we show that it is possible to treat $f(R)=R\pm\mu^4/R$
gravity in a perturbative fashion such that it shows no instabilities
on cosmological scales and, in the solar system, is consistent with
measurements of the PPN parameters. We show that such a theory
produces a spatially flat, accelerating universe, even in the absence
of dark energy and when the matter density is too small to close the
universe in the general relativistic case.
\end{abstract}

\maketitle

Since it was first proposed as an explanation for the present period
of cosmological acceleration~\cite{cdtt04}, the idea of modifying the
Einstein-Hilbert action of General Relativity with new functions of
the Ricci curvature $R$ has been the subject of a great deal of
attention. In particular, its effect on the homogeneous cosmological
expansion, on the growth and evolution of perturbations in a
cosmological scenario, and on precision tests in the solar system,
have all been investigated~\cite[and refs. therein]{hs07}.

To date, all of these investigations acknowledge that $f(R)$ gravity,
considered as a complete theory, faces some serious problems arising
from the set of extra degrees of freedom.  For example, the addition
of an $1/R$ term to the Einstein-Hilbert action is inconsistent with
the first post-Newtonian correction to the metric outside a
star~\cite{esk06}, and also generates unstable solutions for
spherically symmetric stars~\cite{s07}. In the originally suggested
formulation, it also leads to serious instabilities in the evolution
of a homogeneous universe~\cite{dk03} and in the growth of
perturbations in the early universe~\cite{shs07}. It is worthwhile to
emphasize that all these instabilities are not related to
Ostrogradski's theorem~\cite{w06}, but are nevertheless directly
associated with the presence of higher derivatives in the Lagrangian
of the gravitational field.

One avenue that can suppress the instabilities of
$1/R$ gravity is to derive the field equations from
from the Langrangian in the so-called Palatini formalism, \emph{i.e.},
by assuming that the connection and the metric are independent
fields~\cite{sl06}. This reduces the order of the field equations and
resolves the instabilities.  

However, it was recently shown that the
field equations that are generated with the Palatini formalism do not
allow for consistent solutions of the metrics of polytropic
stars~\cite{bs07}. As a result, this approach cannot be considered
viable for gravity theories with non-linear Lagrangian actions. While
there are many, more elaborate, functions of $R$ that may produce
viable theories~\cite{no06,no07}, $1/R$ gravity appears in general to
suffer from the pathological degrees of freedom.

Matters change, however, if we consider a polynomial $f(R)$ Lagrangian
to be a truncation of a perturbative expansion of a more general theory. This, as we shall see, can
radically change the properties of the system. There are two general
ways that can lead to a perturbative expansion in power of the
curvature.

In one case, the $f(R)$ gravity can be simply a Taylor (or Laurent)
expansion of a non-polynomial Lagrangian. It is often the case
that the naive pathologies of the field theory do not appear in
the more general actions. there exists a large class of theories whose perturbative approximations produce additional, fictional degrees of freedom when treated naively -- including not only the semi-classical limit of Quantum Electrodynamics, but also the computation of String Theory gravitational corrections to General Relativity~\cite{s90b}.

In these cases, while the fundamental theory is second-order and local, the effective field theory is higher-order and will generically lead to unstable, and even fictional, degrees of freedom even in regimes where the perturbative quantities are small~\cite{s90a}.

In a different situation, $f(R)$ gravity is an expansion of a fundamental theory that is of second-order but
non-local. In this case, the expansion generically introduces
fictional degrees of freedom even in regimes
where the perturbative quantities are small. Such pathologies,
however, can be removed in the same, mathematically rigorous, and
consistent fashion to restore the correct behavior of the
theory~\cite{ew89}.

In the end, however, the origin of the justification does not matter for the actual results of the analysis which are the same in each case; we may simply consider $1/R$ gravity to be a perturbative approximation (``perturbative $f(R)$'') instead of considering it an
exact theory (``exact $f(R)$.'') This fundamentally
changes the analysis of various cosmological and astrophysical
phenomena. This interpretation removes the additional degrees of
freedom from the field equations and, as we show below, it cures the
theory from classical instabilities. In addition, the perturbative and
solar system properties are radically changed.

It is important to emphasize here that while our approach shares some (but not all)
of the mathematical methods associated with effective field theories in the quantum realm, 
the reasoning behind the constraints imposed upon the apparent degrees of freedom is not as restricted. 

In effective field theories, there is a high-energy (``ultraviolet'') limit above
which we expect new physics to appear -- allowing us to freeze out
degrees of freedom that have mass scales
above this limit because conservation of energy limits their excitation. 

Here, by contrast, the mechanism that prevents the excitation is left unspecified; the
techniques of the analysis allow us to produce a consistent answer in the perturbative
regime. Such techniques can be applied not only to constraints from energy conservation but also, for example, to non-local theories where the constraint is implicit, or to cases where non-canonical kinetic terms constrain perturbative degrees of freedom to track a cosmological background~\cite{acg07}. We need assume only that such a constraint exists, and how it appears to leading order in some limit, to ask its perturbative consequences.

In the first section, we introduce the main techniques of our
analysis, and demonstrate that, even within our perturbative approach,
the $1/R$ term in the action leads to late-time cosmic
acceleration. We discuss the limitations of the approach for providing
definite predictions of the homogeneous expansion.

In the second section we examine the behavior of cosmological
perturbations to demonstrate that there are no instabilities for the
theory in the matter dominated regime. In the third section we show
that perturbative $f(R)$ gravity can pass solar system tests at the current
level of experimental sensitivity. Finally, we summarize our results
and discuss the provocative ways they reflect upon studies of $f(R)$ gravity.

\section{Homogeneous Expansion to First Order.} 

A great variety of functions $f(R)$ for the Lagrangian of the
gravitational field have been proposed. The original suggestion,
$f(R)$ taken to be $R-\mu^4/R$ (the ``CDTT'' case), suffers a number
of problems under the exact paradigm. However, as we show here, under
the perturbative analysis it survives very well, and we shall consider
it in detail.

Assuming an expanding universe with signature $(-,+,+,+)$ and spatial
curvature $k$, we first compute the ``exact'' equation of motion for
$R+\mu^4/R$ gravity (for convenience, we flip the sign -- CDTT can be
considered to have a negative value for $\mu^4$.) We find, in
accordance with the literature,
\begin{eqnarray*}
\left(1-\frac{\mu^4}{R^2}\right)R_{\mu\nu}-\frac{1}{2}\left(1+\frac{\mu^4}{R^2}\right)Rg_{\mu\nu} & & \\
-\mu^4[g_{\mu\nu}\nabla_{\alpha}\nabla^{\alpha}-\nabla_{(\mu}\nabla_{\nu)}]R^{-2} & = & 8\pi GT_{\mu\nu}\;.
\end{eqnarray*}
There are two independent components in the Einstein equation; we can
take the time-time and trace components. Respectively,
\begin{eqnarray}
\frac{R}{2}-\frac{3\ddot{a}}{a}+\mu^4\left(\frac{1}{2R}+\frac{6\dot{a}\dot{R}}{aR^3}+\frac{3\ddot{a}}{aR^2}\right) & = & 8\pi G\rho, \label{tt} \\
R+\mu^4\left(\frac{3}{R}+\frac{18\dot{a}\dot{R}}{aR^3}-\frac{18\dot{R}^2}{R^4}+\frac{6\ddot{R}}{R^3}\right) & = & -8\pi G\rho, \label{trace}
\end{eqnarray}
where $R$, the Ricci scalar, is
\begin{equation}
R=\frac{6(\dot{a}^2+a\ddot{a}+k)}{a^2},
\end{equation}
and overdots are derivatives with respect to time. 

Whatever the matter-energy content of the Universe, since $R$ contains
second derivatives of $a$, this set of equations is fourth order and
there are thus two extra degrees of freedom. Here we have, for simplicity, assumed that all matter is pressureless dust -- reasonable for large scales in the ``matter dominated'' regime.

In most work on $f(R)$ gravity, the equations are rewritten so that
these extra degrees are absorbed into a scalar $\phi$ field that is governed
by a second order equation of motion.  We wish, on the other hand, to
consider $\mu^4/R$ as only the first term in a series expansion. We
remain agnostic about what the next term looks like, as indeed we are
allowed to do in the context of a perturbative expansion.

We will thus seek a solution to Eqs.~\ref{tt} and ~\ref{trace} valid
only to $\mathcal{O}(\mu^4)$. To do so is simple. For the terms above
multiplied by $\mu^4$, we insert the zeroth order solutions --
\emph{i.e.}, the solutions to the ordinary Friedman equation. 

Doing so, and using the fact that the stress-energy tensor is
conserved even in $f(R)$ gravity~\cite{w84}, we find the new, modified
Friedman equation to be
\begin{equation}
H^2=\frac{8\pi G\rho}{3}-\frac{k}{a^2}-\frac{3\mu^4}{32\pi^2 G^2\rho^2}\left(\frac{k}{a^2}-\frac{8\pi G\rho}{3}\right)+\cdots,
\label{modf}
\end{equation}
where the dots remind us that this approximation is good only to first
order. 

From this result, we see that a standard matter-dominated era exists
in a perturbative $1/R$ theory; this is in direct contrast to the
behavior in exact $1/R$~\cite{a07,f06,hs07}, where the usual $t^{2/3}$
scaling of the matter era is unstable to a so-called ``$\phi MDM$ era'', where the scaling is now
$t^{1/2}$. In both theories the matter dominated solution exists; the
suppression of the extra degrees of freedom makes it stable in
perturbative $1/R$.

\begin{figure}
\includegraphics[width=3.275in]{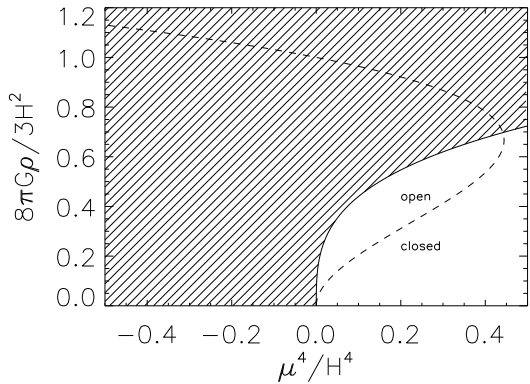}
\caption{The phase space of perturbative $1/R$ gravity. The value 
of $\mu$, referenced to the Hubble parameter, is on the horizontal
axis and the matter density is on the vertical axis. The unshaded
region is where one finds acceleration, and the dashed line is the
locus of points for which the spatial curvature is zero. This graph
shows that when $(\mu/H)^4\simeq -0.2$, cosmic
acceleration {\em and\/} a spatially flat universe are consistent with
the amount of matter in the universe inferred by traditional methods.}
\label{niceplot}
\end{figure}

By inspection of Eq.~\ref{modf}, one can see that in perturbative
$1/R$ it is possible to have a spatially flat universe even if the
matter density appears insufficient in the General Relativistic
case. Further, one can see that many choices for $k$ and $\mu^4$ will
lead to accelerated expansion. Fig.~\ref{niceplot} shows the various
regimes, for different values of $\mu^4$ and the matter density, the
latter phrased as the ``classical'' GR quantity $8\pi G\rho/3H^2$,
equal to $\Omega_m$ in the $\mu$ equal to zero case. Indeed, for
$(\mu/H)^4\simeq 0.2$, a classical GR value of $\Omega_{\rm m}\simeq
0.3$ leads to both a spatially flat universe and to late-time cosmic
acceleration. It is important to recognize that if the ``control parameter'' of the
perturbative expansion is taken to be $(\mu/H)^4$, this solution approaches the limit in which one expects
results to be reliable; the extent to which the onset of ``lambda domination'' allows for 
perturbative approaches requires additional study.

Acceleration in our model begins to appear at high curvatures where
the perturbative expansion is valid. However, interpreting this plot
requires a degree of caution: in particular, it must be emphasized
that the results here are good only to $\mathcal{O}(\mu^4)$; whether
or not we can reproduce, e.g., the currently measured deceleration
parameter at refshift zero, would require knowledge of the full theory
of which $1/R$ is only an approximation.

When the relevant physical quantities -- either the curvature squared,
$R^2$, or the matter density, $(G\rho)^2$ -- become of order $\mu^4$,
the perturbative analysis here can no longer be trusted and the
complete field equation needs to be solved to generate predictions. By
construction, the present day universe is in a regime where these
assumptions break down and, hence, quantitative predictions of this
theory for comparison with observations will require specification of
the theory beyond the order considered here.

Note also that, as expected, the two extra degrees of freedom have
disappeared; simply specifying the scale factor and its derivative at
some early time is sufficient to determine the complete future
evolution of the system. The phenomenon of a finite-term approximation
to an infinite series in an action or equation of motion leading, in a
naieve treatment, to extraneous degrees of freedom, is well known. Our
manipulations above are a canonical solution to the
problem~\cite{ew89}.

\section{Cosmological Perturbations.} 

As is well known, the CDTT case, considered as an exact theory, is
unstable to perturbations on cosmological scales. This can be seen by
an examination of CDTT gravity in the Einstein frame where it is found
that the extra degrees of freedom -- considered now as a scalar field
-- show an instability on superhorizon scales that grows larger at
earlier times as the curvature increases~\cite{sh07}.

While the opposite sign, \emph{i.e.}, the action $R+\mu^4/R$, is
stable even in the exact case, and Fig.~\ref{niceplot} shows that it
is this choice that leads to acceleration at first order, it is worth
noting that the stability of the evolution does not depend on the sign
of $\mu^4$ and instead is related to the disappearance of these extra
degrees of freedom.

This can be seen, trivially, by examining the perturbative Friedman
equation, Eq.~\ref{modf}. At early times, when the matter density
$\rho$ is much larger than $\mu^4$, we recover the classical Einstein
solution and perturbations in the curvature grow in the usual
Einstein-de Sitter fashion. The growth occurs on timescales of order
$1/H$ and not, as in the case of the ``exact'' theory, on the much
shorter timescale of order $\mu^2/H^3$.

The stability of our solution is related to the absence of additional
degrees of freedom that would allow one to relax into a high-density
but low-curvature regime. As we can see from Eq.~\ref{trace}, the
deviation from the classical curvature-density relation is constrained
to be of $\mathcal{O}(\mu^4)$ or smaller.

\section{Solar System Tests.}\label{iii} 

We finally consider solar system constraints. In particular, we
examine whether or not tight constraints on the Parametrized
Post-Newtonian (PPN) parameters can rule out $1/R$ gravity considered
as a perturbative expansion. As suggested by Ref.~\cite{cse07}, and
reasserted by Ref.~\cite{esk06}, non-perturbative $1/R$ gravity is
equivalent to a scalar-tensor theory that may have already been ruled out
using solar system tests.

Our analysis of the problem will be conceptually similar to that of
our cosmological investigation; we note that are results are in
agreement with a ``post-Newtonian'' (\emph{i.e.}, weak field) analysis
under a similar ``perturbative'' interpretation of
$1/R$~\cite{scw06}. We will at all times make our approximations to
$\mathcal{O}(\mu^4)$, and test the assumption that the solution stays
perturbatively close to that found for General Relativity. Since our
perturbative expansion can not handle the vacuum regime, we will
assume that $(G\rho)^2$ always remains much larger than $\mu^4$ -- a
very reasonable assumption given the ambient densities of the solar
system.

The relevant equation of motion is the trace equation~\cite{esk06},
written here without reference to a particular background spacetime as
\begin{equation}
\label{tracefree}
\Box\frac{\mu^4}{R^2}-\frac{R}{3}-\frac{\mu^4}{R}=\frac{8\pi GT}{3}.
\end{equation}
Inside the star -- in the presence of matter -- the background Ricci
curvature that we perturb around is $8\pi G\rho$. We define $c(r)$ to
be the (dimensionful) departure from the background solution:
\begin{equation}
c(r)\equiv R^2(r)-(8\pi G\rho)^2~~~ [=\mathcal{O}(\mu^4)\textrm{ or higher}].
\end{equation}
We then rewrite Eq.~\ref{tracefree} in terms of $c(r)$ and $\rho(r)$, finding
\begin{eqnarray}
-3(8\pi G\rho)^2\mu^4\left[\frac{(8\pi G\rho)^5r}{2}-rc^{\prime 2}-(16\pi G)^2r\rho\rho^\prime c^\prime\right. \nonumber \\
+\frac{1}{2}(8\pi G\rho)^2(2c^\prime-6(8\pi G)^2r\rho^{\prime 2}+rc^{\prime\prime}) \nonumber \\
\left.\frac{{}^{}}{}+(8\pi G)^4\rho^3(2\rho^\prime+r\rho^{\prime\prime})\right] - c\left(\frac{1}{2}(8\pi G\rho)^7r \right. \nonumber \\
-\mu^4\left[\frac{3}{4}(8\pi G\rho)^5r-9rc^{\prime 2}-36(8\pi G)^2 r\rho c^\prime\rho^\prime\right. \nonumber \\
3(8\pi G\rho)^2(10(8\pi G)^2r\rho^{\prime 2}-2c^\prime-rc^{\prime\prime}) \nonumber \\
\left.\left.\frac{{}^{}}{}+6(8\pi G)^4\rho^3(2\rho^\prime+r\rho^{\prime\prime})\right]\right) = 0,
\label{large}
\end{eqnarray}
where primes denote derivatives with respect to radius. If we assume that the first and second derivatives of $c(r)$ are of order $c(r)/r$ and $c(r)/r^2$, we can solve Eq.~\ref{large} for $c(r)$ self-consistently to
$\mathcal{O}(\mu^4)$. We find
\begin{equation}
c(r) = -\mu^4\left(6+\frac{3\rho^\prime}{G\pi r \rho^2}-\frac{9\rho^{\prime 2}}{2G\pi\rho^3}+\frac{3\rho^{\prime\prime}}{2G\pi\rho^2}\right),
\end{equation}
This expression is finite for physical boundary conditions at the origin. Note
that the coefficient of $\mu^4$ here may become large depending on the
system; however, if $\mu^4$ is chosen to be the cosmic acceleration
scale, the dimensionless departure from General Relativity --
\emph{i.e.}, $c(r)/(G\rho)^2$ -- is of order
\begin{equation}
\label{consistency}
\alpha=\frac{r^6_{\mathrm{curv}}}{r^2_{\mathrm{object}}r^4_{\mathrm{horizon}}},
\end{equation}
where $r_\mathrm{horizon}$ is the Hubble distance, $r_\mathrm{object}$
is the size of the object, and $r_\mathrm{curv}$ its (gravitational)
radius of curvature, $c/\sqrt{G\rho}$. For a laboratory experiment,
$\alpha$ is $10^{-32}$, for the Sun, $10^{-50}$, for the inner solar
system, $10^{-35}$, and for the galaxy $10^{-13}$ -- easily satisfying
the current PPN bounds. In all these cases the assumption made in the
course of the solution about the derivatives of $c(r)$ is consistent,
as can be seen by examination.

We note that our solution here is conceptually distinct from the debates as to the viability of exact $f(R)$ gravity in the solar system. It has been suggested~\cite{f06,hs07} that a non-cosmological ambient density in the solar system might allow us, through a ``chameleon'' effect, to avoid the no-go analysis of Ref.~\cite{esk06}. In those analyses, the mass of the extra degrees of freedom becomes large enough to ``freeze out'' as in a standard effective field theory. Our analysis here does also rely on a non-zero ambient density, but for an entirely different reason -- to satisfy the validity of a perturbative approximation to a more fundamental theory. We reach equivalent results in the solar system, but not on cosmological scales.

\section{Conclusions.} In this work, we have examined
$f(R)$ gravity in a perturbative fashion and considered its
effects in cosmological and solar-system observations. Instead of
considering it an exact theory, whose higher-order derivatives
demand the introduction of new degrees of freedom, we require it to be
perturbatively close to GR. Such an approach removes the extra degrees of freedom and greatly simplifies
the theory without making it trivially equivalent to GR.

We have demonstrated a number of facets of the theory. We have shown that it avoids instabilities that exist in $f(R)$ gravities on cosmological scales, and also that it can, to leading order, induce acceleration. Further we have shown that in the solar system the theory is consistent with current measurements of the PPN parameters.

The use of our work, at the intersection of observational cosmology
and fundamental physics, is twofold. 

First, we draw attention to the
fact that many common modifications to gravity lead to radically
different results if viewed as perturbative approximations to a more fundamental theory. In several cases, the approach resolves several
traits of the theories that would be, otherwise, prohibitive. 

Second,
we provide an encouraging note for possible experimental verification
of the presence of non-linear terms in the action of the gravitational
field. Broadly speaking, modifications to gravity that look, in
approximation, like the kinds of $f(R)$ expansions we address here,
may be considered \emph{prima facie} reasonable despite the mounting
evidence against their being the exact deviations from the
Einstein-Hilbert action.

\section{Acknowledgments}. S.D. thanks Wayne Hu and D.P thanks 
Bira van Kolck and Sean Fleming for important conversations; both
thank Matias Zaldarriaga, Cai Rong-Gen and Alexei Starobinsky for
important correspondence.

\end{document}